# MicroRNAs preferentially target the genes with high transcriptional regulation complexity

Qinghua Cui[a+], Zhenbao Yu[a+], Youlian Pan[b], Enrico O. Purisima[a] and Edwin Wang[a]*

[a]Biotechnology Research Institute, National Research Council Canada, Montreal, Quebec, H4P 2R2, Canada

[b]Institute for Information Technology, National Research Council Canada, Ottawa, Ontario K1A 0R6, Canada

[+]These authors contributed equally to this work

[*]Correspondence author. Fax: 1-514-496-5143

Email address: edwin.wang@cnrc-nrc.gc.ca (E. Wang)

**Abstract**

Over the past few years, microRNAs (miRNAs) have emerged as a new prominent class of gene regulatory factors that negatively regulate expression of approximately one-third of the genes in animal genomes at post-transcriptional level. However, it is still unclear why some genes are regulated by miRNAs but others are not, i.e. what principles govern miRNA regulation in animal genomes. In this study, we systematically analyzed the relationship between transcription factors (TFs) and miRNAs in gene regulation. We found that the genes with more TF-binding sites have a higher probability of being targeted by miRNAs and have more miRNA-binding sites on average. This observation reveals that the genes with higher *cis*-regulation complexity are more coordinately regulated by TFs at the transcriptional level and by miRNAs at the post-transcriptional level. This is a potentially novel discovery of mechanism for coordinated regulation of gene expression. Gene Ontology analysis further demonstrated that such coordinated regulation is more popular in the developmental genes.

*Key words:* MicroRNA; Transcription factor; Gene expression regulation; MicroRNA target; Transcription factor-binding site

*Abbreviations*: GO, gene ontology; miRNA, microRNA; TF, transcription factor; TFBS, TF-binding site

Gene expression is largely regulated by action of *trans*-factors on the *cis*-elements aligning on the regulatory regions of the genes. Among these *trans*-factors and the *cis*-elements, transcription factors (TFs) and their binding sites (TFBSs) play the most important role in gene expression regulation. Recently, another group of molecules, namely microRNAs (miRNAs), have been found to regulate gene expression at the post-transcriptional level and translational level through base-pairing with target messenger RNAs (mRNAs).

MicroRNAs are noncoding RNAs of ~22-nucleotide in length that are encoded in the chromosomal DNA and transcribed as longer stem-loop-like precursors [1; 2]. Upon transcription, the miRNA precursors are converted to mature miRNA duplex through sequential processing by RNaseIII family of endonucleases Drosha and Dicer [3; 4]. One strand of the processed duplex is incorporated into a silencing complex and guided to target mRNA sequences by base-pairing, resulting in the cleavage of target mRNAs or repression of their productive translation [5; 6]. Roughly 1% of the genes in each respective animal genome are miRNAs [7]. A growing body of evidence has revealed that miRNAs are involved in a variety of biological processes [1; 2] and pathological processes [8], such as embryonic development, cell proliferation, cell differentiation, apoptosis, insulin secretion and carcinogenesis.

Recent computational studies indicate that approximately one third of human genes are potentially regulated by miRNAs and each miRNA on average could target more than 200 genes [7]. Similar to TFs and their regulated genes, miRNAs and their targets appear to form a complex regulatory network. However, it remains unclear whether the gene expression regulation by miRNAs at the post-transcriptional level is coordinated with that by TFs at the transcriptional level. In this study, we systematically analyzed the relationship between the abundance of the TFBSs and miRNA targeting sites of each gene in human genome. We found a positive correlation between these two groups of transcriptional regulators.

**Materials and methods**

*Datasets used in this study.* A dataset representing three transcription factors, OCT4, NANOG and SOX2 and their target genes in human embryonic stem cells were obtained from Boyer et al. [9]. The regulatory relationships of the three transcription factors and their target genes are listed in Supplementary Text File S1.

The genome-wide computationally predicted human miRNA target genes were obtained from Krek et al. [10]. There were a total of 6,243 genes regulated by 168 miRNAs. The miRNAs and their targets are listed in Supplementary Text File S2.

We obtained two TFBS datasets, one from Cora et al. [11] and the other from Xie et al. [12]. We counted the number of TFBS for each gene from these two datasets and presented them in Supplementary Text Files S3 and S4, respectively.

The original Gene Ontology (GO) term file was downloaded from NCBI (ftp://ftp.ncbi.nlm.nih.gov/gene/DATA/). Human GO terms were extracted and listed in Supplementary Text File S5.

*Analysis of the relationship between miRNA target rate and TFBS number of human genes.* We define the miRNA target rate as a ratio of the number of genes that are miRNA targets to the total number of genes in a dataset. To examine the relationship between miRNAs and TFs in gene regulation, the 6,243 human miRNA target genes reported by Krek et al. [10] were mapped onto 9,348 human genes whose numbers of corresponding TFBSs were determined by Cora et al. [11]. We call the number of TFBSs associated with a gene its TFBS-count. We divided these 9,348 human genes into two groups. One group contains miRNA targets (Supplementary Text File S6) and the other group contains the rest of the genes (Supplementary Text File S7). We calculated the average TFBS-count of the genes in each group. A Wilcoxon Ranksum test was performed to determine whether the average number of TFBSs was significantly different between the two groups.

In order to investigate the relationship between the miRNA target rate and the TFBS-count of genes in more details, we calculated the number of TFBSs for each gene in the dataset of Cora et al [11] (Supplementary Text File S3) and grouped the genes according to their TFBS-count. The genes in each group have the same number of TFBSs. Since some groups contain limited numbers of genes, we set a threshold of 100 genes in each group and regrouped the genes if a group contains less than 100 genes. For example, when a group contains N genes (N<100), we randomly selected 100-N genes from the adjacent group that contains the genes with one less TFBS-count. If the number of the genes in the adjacent group (M) is not enough to make a new group (M+N<100), we continued this procedure until the number of genes reaches to the threshold. This ensures that each group has at least 100 genes. Then, we calculated the miRNA target rate for each gene group (Supplementary Text File S8). We performed a similar analysis based on the genes' TFBS information from the dataset of Xie et al. [12]. (Supplementary Text File S9).

*Analysis of the relationship between the number of TFBSs and the number of miRNAs that regulate the same gene.* As many genes are regulated by several different TFBSs and different miRNAs, we analyzed the relationship between the number of TFBSs and the number of miRNAs which target the same genes. To do so, we divided the genes in Supplementary Text File S3 into subgroups based on the number of miRNAs that regulate the genes. The grouping information and genes in each group are listed in Supplementary Text File S10. For example, Group one contains the genes that are not regulated by miRNAs at all, Group two contains the genes that are regulated by one miRNA, and so on. The average number of TFBSs in each group was calculated using a similar method as described above and listed in Supplementary Text File S11. A similar analysis was conducted using the genes' TFBS information from Xie et al [12]. (Supplementary Text File S12).

*Analysis of the relationship between the number of TFBSs and miRNA-target of the genes in Gene Ontology categories.* To investigate which biological processes and functional categories are more regulated by TFs and miRNAs, we assembled genes into functional groups based on the Gene Ontology (GO) annotations. Among the 9,348 genes in the dataset of Cora et al. [11], 6,994 were found to have GO terms. For statistical analysis, we only retained the GO term groups (233 terms) that contain more than 50 genes. To gain an insight into the relationship of gene regulation between TFs and miRNAs in each GO category, we determined the Pearson's correlation between the TFBS-counts and the miRNA target rates within such GO groups with respect to the data distribution among the 233 GO terms.

All intermediate results and data files are accessible at: http://www.bri.nrc.ca/wang/mirna3.html

**Results and discussion**

To study how miRNAs and TFs coordinately regulate genes in human genome, we first took a dataset which represents the regulatory relations between the three TFs, OCT4, NANOG and SOX2 and their target genes in human embryonic stem cells [9]. The regulatory relationships between the three TFs (OCT4, NANOG and SOX2) and their target genes were determined through ChIP-chip method (chromatin immunoprecipitation coupled with DNA microarray) [9]. The three TFs totally regulate 2,046 genes. Among these 2046 genes including the three TFs, 341 are regulated by only one of the three TFs, 1,314 are co-regulated by two of the TFs and 391 are co-regulated by all of the TFs (Table 1). Meanwhile, we obtained a set of computationally predicted human miRNA target genes from Krek et al. [10]. The current miRNA target prediction methods [10; 13-18] are mainly based on the principle of miRNA-target interactions [19], and the accuracy of these methods has been confirmed by experimental validation of randomly selected miRNA targets [20] and by large-scale gene expression profiling studies [21; 22]. Up to 90% of the randomly selected miRNA targets from the predictions by Krek et al. [10] have been validated as true targets [20]. Accordingly, the predicted miRNA targets have been used for genome-wide analysis of miRNA target expression [19; 23-25] and cellular signaling network regulation [26]. We mapped the miRNA targets onto the target genes of the three TFs. We then divided the genes into three groups, in which they are regulated by one, two and all three of the TFs, respectively, and counted the number of genes that are miRNA targets and the number of genes that are not miRNA targets, respectively, in each group. We found that the miRNA targets are enriched in the group of genes targeted by more TFs (Table 1). Fisher exact test confirmed the significance of the observation (P = 0.031). This result indicates that a gene that is regulated by more TFs is also more likely to be a target of miRNAs.

A gene is usually regulated by more than three TFs. To validate and expand above observation, we examined the relationship between TFs and miRNAs for gene regulation in a genome-wide scale. In the past few years, genome-wide identification of TFBSs has been extensively studied by using various bioinformatics methods. Among these methods, the comparative genomics approach emerges as one of the most effective and relatively accurate approaches for identifying potential *cis*-regulatory elements in genomes [27; 28]. *Cis*-regulatory elements have been used to study gene expression divergence [29-32] and tissue-specific gene expression profiles [12; 33]. TFBSs have been also used to reconstruct gene regulatory networks, find co-regulated genes and infer evolutionary insights [34-37]. One report found that 98% of known TFBSs of skeletal muscle-specific TFs are confined to 19% of the most conserved sequences between human and mouse [38]. Furthermore, several integrative methods that combine phylogenetic footprinting with others such as gene microarray and GO have been developed to filter out false positives in the TFBS identification [11]. TFBSs, or *cis*-regulatory elements are normally located in the promoter region of a gene. TFs regulate a gene through binding to the TFBSs of the gene. Basically, the more TFBSs a gene has, the more complex its regulation can be as provided by various possible combinations of TFs. We took two datasets of putative human TFBSs identified through comparative genomics [11; 12]. In the first dataset [11], each gene on average contains 20 putative TFBSs in its promoter region. After mapping miRNA targets reported in [10] onto the 9,348 genes, we found that 42.16% of them are miRNA targets and the average TFBS-count of the miRNA target genes is significantly higher than that of the non-target genes (24.2 vs. 17.4, $P < 1.9 \times 10^{-55}$, Wilcoxon Ranksum test, Table 2). A similar result (86.3 vs. 67.1, Wilcoxon Ranksum test, $P < 1.5 \times 10^{-216}$) was obtained based on the other TFBS dataset [12].

More detailed analysis was performed by grouping these genes based on the TFBS-count and calculating miRNA target rate in each group. As shown in Fig. 1A, the TFBS-count is significantly correlated with the miRNA target rate (Pearson's correlation coefficient $r = 0.9432$, $P < 3.5 \times 10^{-68}$). For example, the miRNA target rate is doubled from the group of genes that have less than 10 TFBSs to those that have more than 100 TFBSs (from ~35% to ~70%). A similar result was obtained using the TFBS dataset from Xie et al. [12] (Fig. 1C, $r = 0.9680$, $P < 3.9 \times 10^{-113}$). These results are consistent with the finding in the human stem cell gene regulation and therefore strongly suggest that miRNAs preferentially target the genes that bear more TFBSs has broad applicability.

Since many genes can be targeted by more than one miRNAs, we analyzed the relationship between the number of miRNAs and the number of TFBSs in the same genes. We found a significant correlation (Fig. 1B, $r = 0.7364$, $P < 6.1 \times 10^{-12}$). A similar result was obtained when using the TFBS dataset of Xie et al. [12] (Fig. 1D, $r = 0.7200$, $P < 9.5 \times 10^{-12}$). These results suggest that the genes that are targeted by more miRNAs have more TFBSs.

Taken together, these results indicate that the complexity of gene regulation by miRNAs at the post-transcriptional level is positively related to the complexity of gene regulation by TFs at the transcriptional level in human genome.

To understand which biological processes and functional categories are more coordinately regulated by both TFs and miRNAs, we assembled the genes from [11] into functional groups based on Gene Ontology (GO) annotation, and calculated average TFBS-count and miRNA target rate in each GO group. As shown in Figure 2, the GO groups that contain genes with a higher average TFBS-counts usually have higher miRNA target rates (Pearson's correlation coefficient r=0.51, $P<2.2\times10^{-16}$). Similarly, the GO groups that contain genes with a lower average TFBS-count also have a lower miRNA target rate. This result suggests that the complexity of gene regulation by TFs and miRNAs are positively associated within most GO functional groups. To determine which functional groups of genes are most complexly regulated by both TFs and miRNAs, we selected the top 14 GO groups (Fig. 2, the up-right corner) that have both the highest average TFBS-counts and high miRNA target rates. We found that many of them are involved in development, such as GO0007275 (development), GO0007399 (nervous system development) and GO0001501 (skeletal development), and development-related processes, such as GO0030154 (cell differentiation), GO0009653 (morphogenesis), and most of the rest are involved in potassium channel and transport, and protein phosphorylation (Fig. 2, right panel). This reflects the fact that gene regulations of development and signal transduction are highly complex, requiring multiple TFBSs and miRNAs.

In conclusion, human genes, especially the genes involved in developments, is coordinately regulated by both TFs at the transcriptional level and miRNAs at the post-transcriptional level. Genes that are more complexly regulated at transcriptional level are more frequently turned on and more differentially expressed at different temporal and spatial conditions and therefore also require to be more frequently turned off. MiRNAs as negative regulators can exert the turning-off function at post-transcriptional level through repressing mRNA translation and/or mediating cleavage of mRNAs. The coordination between TFs and miRNAs in gene expression regulation found in this research reveals a potential mechanism of gene regulation in human genome. We also found that such coordinately regulated genes are enriched in certain GO functions, particularly in those involved in developmental processes.

**Acknowledgements**

This work was partially supported by Genomics and Health Initiative, National Research Council Canada.

Reference List

[1] D.P. Bartel, MicroRNAs: genomics, biogenesis, mechanism, and function, Cell 116 (2004) 281-297.

[2] V. Ambros, The functions of animal microRNAs, Nature 431 (2004) 350-355.

[3] V.N. Kim, MicroRNA biogenesis: coordinated cropping and dicing, Nat. Rev. Mol. Cell Biol. 6 (2005) 376-385.

[4] B.R. Cullen, Transcription and processing of human microRNA precursors, Mol. Cell 16 (2004) 861-865.

[5] M.A. Carmell, G.J. Hannon, RNase III enzymes and the initiation of gene silencing, Nat. Struct. Mol. Biol. 11 (2004) 214-218.

[6] G. Meister, T. Tuschl, Mechanisms of gene silencing by double-stranded RNA, Nature 431 (2004) 343-349.

[7] S. Griffiths-Jones, S. Moxon, M. Marshall, A. Khanna, S.R. Eddy, A. Bateman, Rfam: annotating non-coding RNAs in complete genomes, Nucleic Acids Res. 33 (2005) D121-D124.

[8] S.M. Hammond, MicroRNAs as oncogenes, Curr. Opin. Genet. Dev. 16 (2006) 4-9.

[9] L.A. Boyer, T.I. Lee, M.F. Cole, S.E. Johnstone, S.S. Levine, J.P. Zucker, M.G. Guenther, R.M. Kumar, H.L. Murray, R.G. Jenner, D.K. Gifford, D.A. Melton, R. Jaenisch, R.A. Young, Core transcriptional regulatory circuitry in human embryonic stem cells, Cell 122, (2005) 947-956.

[10] A. Krek, D. Grun, M.N. Poy, R. Wolf, L. Rosenberg, E.J. Epstein, P. MacMenamin, P. da, I, K.C. Gunsalus, M. Stoffel, N. Rajewsky, Combinatorial microRNA target predictions, Nat. Genet. 37 (2005) 495-500.

[11] D. Cora, C. Herrmann, C. Dieterich, C.F. Di, P. Provero, M. Caselle, Ab initio identification of putative human transcription factor binding sites by comparative genomics, BMC. Bioinformatics. 6 (2005) 110.

[12] X. Xie, J. Lu, E.J. Kulbokas, T.R. Golub, V. Mootha, K. Lindblad-Toh, E.S. Lander, M. Kellis, Systematic discovery of regulatory motifs in human promoters and 3' UTRs by comparison of several mammals, Nature 434 (2005) 338-345.

[13] A.J. Enright, B. John, U. Gaul, T. Tuschl, C. Sander, D.S. Marks, MicroRNA targets in Drosophila, Genome Biol. 5 (2003) R1.

[14] B. John, A.J. Enright, A. Aravin, T. Tuschl, C. Sander, D.S. Marks, Human MicroRNA targets, PLoS. Biol. 2 (2004) e363.

[15] M. Kiriakidou, P.T. Nelson, A. Kouranov, P. Fitziev, C. Bouyioukos, Z. Mourelatos, A. Hatzigeorgiou, A combined computational-experimental approach predicts human microRNA targets, Genes Dev. 18 (2004) 1165-1178.

[16] B.P. Lewis, C.B. Burge, D.P. Bartel, Conserved seed pairing, often flanked by adenosines, indicates that thousands of human genes are microRNA targets, Cell 120 (2005) 15-20.

[17] A. Stark, J. Brennecke, R.B. Russell, S.M. Cohen, Identification of Drosophila MicroRNA targets, PLoS. Biol. 1 (2003) E60.

[18] A. Stark, J. Brennecke, N. Bushati, R.B. Russell, S.M. Cohen, Animal MicroRNAs confer robustness to gene expression and have a significant impact on 3'UTR evolution, Cell 123 (2005) 1133-1146.

[19] J. Brennecke, A. Stark, R.B. Russell, S.M. Cohen, Principles of microRNA-target recognition, PLoS. Biol. 3 (2005) e85.

[20] N. Rajewsky, microRNA target predictions in animals, Nat. Genet. 38 Suppl (2006) S8-13.

[21] J. Krutzfeldt, N. Rajewsky, R. Braich, K.G. Rajeev, T. Tuschl, M. Manoharan, M. Stoffel, Silencing of microRNAs in vivo with 'antagomirs', Nature 438 (2005) 685-689.

[22] L.P. Lim, N.C. Lau, P. Garrett-Engele, A. Grimson, J.M. Schelter, J. Castle, D.P. Bartel, P.S. Linsley, J.M. Johnson, Microarray analysis shows that some microRNAs downregulate large numbers of target mRNAs, Nature 433 (2005) 769-773.

[23] K.K. Farh, A. Grimson, C. Jan, B.P. Lewis, W.K. Johnston, L.P. Lim, C.B. Burge, D.P. Bartel, The widespread impact of mammalian MicroRNAs on mRNA repression and evolution, Science 310 (2005) 1817-1821.

[24] P. Sood, A. Krek, M. Zavolan, G. Macino, N. Rajewsky, Cell-type-specific signatures of microRNAs on target mRNA expression, Proc. Natl. Acad. Sci. U.S.A 103 (2006) 2746-2751.

[25] Z. Yu, Z. Jian, S.H. Shen, E. Purisima, E. Wang, Global analysis of microRNA target gene expression reveals that miRNA targets are lower expressed in mature mouse and Drosophila tissues than in the embryos, Nucleic Acids Res. (2006) in press.

[26] Q. Cui, Z. Yu, E.O. Purisima, E. Wang, Principles of microRNA regulation of a human cellular signaling network, Mol. Syst. Biol. 2 (2006) 46.

[27] C.T. Harbison, D.B. Gordon, T.I. Lee, N.J. Rinaldi, K.D. Macisaac, T.W. Danford, N.M. Hannett, J.B. Tagne, D.B. Reynolds, J. Yoo, E.G. Jennings, J. Zeitlinger, D.K. Pokholok, M. Kellis, P.A. Rolfe, K.T. Takusagawa, E.S. Lander, D.K. Gifford, E. Fraenkel, R.A. Young, Transcriptional regulatory code of a eukaryotic genome, Nature 431 (2004) 99-104.

[28] T. Wang, G.D. Stormo, Identifying the conserved network of cis-regulatory sites of a eukaryotic genome, Proc. Natl. Acad. Sci. U.S.A 102 (2005) 17400-17405.

[29] C.I. Castillo-Davis, D.L. Hartl, G. Achaz, cis-Regulatory and protein evolution in orthologous and duplicate genes, Genome Res. 14 (2004) 1530-1536.

[30] G. Haberer, T. Hindemitt, B.C. Meyers, K.F. Mayer, Transcriptional similarities, dissimilarities, and conservation of cis-elements in duplicated genes of Arabidopsis, Plant Physiol 136 (2004) 3009-3022.

[31] W.H. Li, J. Yang, X. Gu, Expression divergence between duplicate genes, Trends Genet. 21 (2005) 602-607.

[32] Z. Zhang, J. Gu, X. Gu, How much expression divergence after yeast gene duplication could be explained by regulatory motif evolution? Trends Genet. 20 (2004) 403-407.

[33] A.D. Smith, P. Sumazin, Z. Xuan, M.Q. Zhang, DNA motifs in human and mouse proximal promoters predict tissue-specific expression, Proc. Natl. Acad. Sci. U.S.A 103 (2006) 6275-6280.

[34] D.A. Rodionov, I.L. Dubchak, A.P. Arkin, E.J. Alm, M.S. Gelfand, Dissimilatory metabolism of nitrogen oxides in bacteria: comparative reconstruction of transcriptional networks, PLoS. Comput. Biol. 1 (2005) e55.

[35] S. Imoto, T. Higuchi, T. Goto, K. Tashiro, S. Kuhara, S. Miyano, Combining microarrays and biological knowledge for estimating gene networks via Bayesian networks, Proc. IEEE Comput. Soc. Bioinform. Conf. 2 (2003) 104-113.

[36] S. Tang, S.L. Tan, S.K. Ramadoss, A.P. Kumar, M.H. Tang, V.B. Bajic, Computational method for discovery of estrogen responsive genes, Nucleic Acids Res. 32 (2004) 6212-6217.

[37] V.X. Jin, G.A. Singer, F.J. gosto-Perez, S. Liyanarachchi, R.V. Davuluri, Genome-wide analysis of core promoter elements from conserved human and mouse orthologous pairs, BMC Bioinformatics 7 (2006) 114.

[38] W.W. Wasserman, M. Palumbo, W. Thompson, J.W. Fickett, C.E. Lawrence, Human-mouse genome comparisons to locate regulatory sites, Nat. Genet. 26 (2000) 225-228.

**Figure legends**

Fig. 1. Correlation between TFs and miRNAs in gene regulation. The analysis was performed using two TFBS datasets from Cora et al. [11] (A and B) and Xie et al. [12] (C and D), respectively, and one miRNA target dataset from Krek et al. [10] (A-D). (A) and (C) Genes were grouped based on the number of TFBSs. MicroRNA targets were mapped onto these genes and miRNA target rate in each group was calculated. (B) and (D) Genes were grouped based to the number of targeting miRNAs. The average number of TFBSs was calculated in each group.

Fig. 2. Correlation between average TFBS-counts and miRNA target rates within functional GO groups. Genes were assembled into functional groups according to the GO annotations. The average TFBS-counts of the genes with such GO groups was calculated using the TFBS information from Cora et al. [11] and the miRNA target rates of the groups were calculated using the miRNA target dataset reported by Krek et al. [10]. All GO groups (233 terms) that contain more than 50 genes were plotted. The GO terms with both relatively high average number of TFBSs and high miRNA target rates are indicated in the up-right corner box.

Table 1

MiRNA target rate in each subgroup of the genes in human embryonic stem cell

| Group* | 1 | 2 | 3 |
|---|---|---|---|
| Number of total genes | 341 | 1314 | 391 |
| Number of miRNA target genes | 141 | 550 | 192 |
| Number of non-miRNA targets | 200 | 764 | 199 |
| P value | | 0.031 | |

*The genes were grouped according to the number of TFs by which they are regulated.

Table 2

Average TFBS-counts of miRNA targets and non-miRNA targets*

| | Total genes | miRNA targets | Non-miRNA targets |
|---|---|---|---|
| Nubmer of genes | 9348 | 3942 | 5406 |
| Average TFBS-count | 20.4 | 24.2 | 17.4 |
| P value** | $1.9 \times 10^{-55}$ | | |

*TFBS data were obtained from Cora et al.'s report [11]

**Wilcoxon Ranksum test